\documentclass{INTERSPEECH2023}

\usepackage{hyperref}       
\usepackage{url}            
\usepackage{booktabs}       
\usepackage{amsfonts}       
\usepackage{nicefrac}   
\usepackage{multirow}
\usepackage{microtype}      
\usepackage{xcolor}         
\usepackage{graphicx}
\usepackage{algorithm}
\usepackage{algorithmic}
\usepackage{pythonhighlight}
\usepackage{mathrsfs}
\usepackage{amsmath}
\usepackage{float}

\interspeechcameraready

\title{Zero-Shot Accent Conversion using Pseudo Siamese Disentanglement Network}
\name{Dongya Jia, Qiao Tian, Kainan Peng, Jiaxin Li, Yuanzhe Chen, Mingbo Ma, Yuping Wang, Yuxuan Wang}
\address{
Speech, Audio \& Music Intelligence (SAMI), ByteDance
}
\email{\{jiadongya,
tianqiao.wave, 
kainan.peng, 
lijiaxin.drany01, 
chenyuanzhe, 
mingbo.ma, wangyuping, 
wangyuxuan.11
\}@bytedance.com}

\begin{document}

\maketitle
 
\begin{abstract}
The goal of accent conversion (AC) is to convert the accent of speech into the target accent while preserving the content and speaker identity. 
AC enables a variety of applications, such as language learning, speech content creation, and data augmentation.
Previous methods rely on reference utterances in the inference phase or are unable to preserve speaker identity.
To address these issues, we propose a zero-shot reference-free accent conversion method, which is able to convert unseen speakers' utterances into a target accent.
Pseudo Siamese Disentanglement Network (PSDN) is proposed to disentangle the accent from the content representation.
Experimental results show that our model generates speech samples \footnote{Some audio samples: \url{https://faceless-rex.github.io/publications/zero-shot-AC-PSDN/}} with much higher accentedness than the input and comparable naturalness, on two-way conversion including foreign-to-native and native-to-foreign. 
\end{abstract}
\noindent\textbf{Index Terms}: Accent Conversion, zero-shot, feature disentanglement

\section{Introduction}

Accent conversion (AC) basically aims to change the accent of speech while retaining the information of content and speaker identity. The applications can be divided into foreign-to-native and native-to-foreign with regard to the selection of the target accent. For foreign-to-native conversion, language learners can use this technique to convert their voice into a native accent while preserving their identity, then they can imitate the converted speech for better learning.
For native-to-foreign conversion, this technique can greatly augment the training set for many speech tasks, such as automatic speech recognition (ASR), by generating various foreign-accent speech. Additionally, in the scenario of video content creation, an accent can be viewed as a special speech style, and thus creators can use AC to make their narration more diverse.


Previous approaches to AC have two major limitations. First, those reference-based methods need
 target-accent utterances as references during inference~\cite{Zhao2018, zhao19f_interspeech, accentron, li2020improving}. This essentially limits the practical use because it is hard to match arbitrary text from inputs. Second, most methods are unable to preserve the speaker identity for unseen speakers~\cite{Zhao2018, zhao19f_interspeech, zhao2021converting, nguyen22d_interspeech, waris2022interspeech}. To address these issues, in the paper, we propose a zero-shot reference-free AC system from the perspective of feature disentanglement. For the task of AC, parallel data, which are utterances uttered by the same speakers with the same content but different accents, barely exist.
In contrast, non-parallel data contain varied but unpaired information and are widely available.
To take full advantage of non-parallel data, it is beneficial to decompose speech into different independent features. In our method, speaker identity, content, and accent are modeled separately.

To represent the content, Bottleneck features (BNFs) are first extracted from a pre-trained automatic speech recognition (ASR) model and then transformed by a content encoder since ASR BNF has been widely used to represent the content in previous works\cite{accentron, zhao2021converting}.
But apart from linguistic information, we find BNFs still contain accent-related information. 
To separate the content and accent, we propose Pseudo Siamese Disentanglement Network (PSDN) to reduce the source accent in the content representation, and then generate the target accent based on the content. For speaker timbre modeling, the speaker timbre feature is extracted by a timbre encoder from the Mel-spectrogram. To improve the generalization ability on unseen speakers, a speaker augmentation method is proposed and applied only at the timbre extraction stage. 

The contributions of this paper include: (1) The proposed method is a zero-shot accent conversion 
method, which is able to convert unseen speakers' utterances into a target accent while greatly preserving the voice identity and content. (2) The proposed method yields high accentedness and naturalness in two-way conversion, including foreign-to-native and native-to-foreign.   (3) Our model is trained on non-parallel data and allows for a limited number of target-accent speakers, such as 1.

\begin{figure*} 
    \centering
    \includegraphics[scale=0.5]{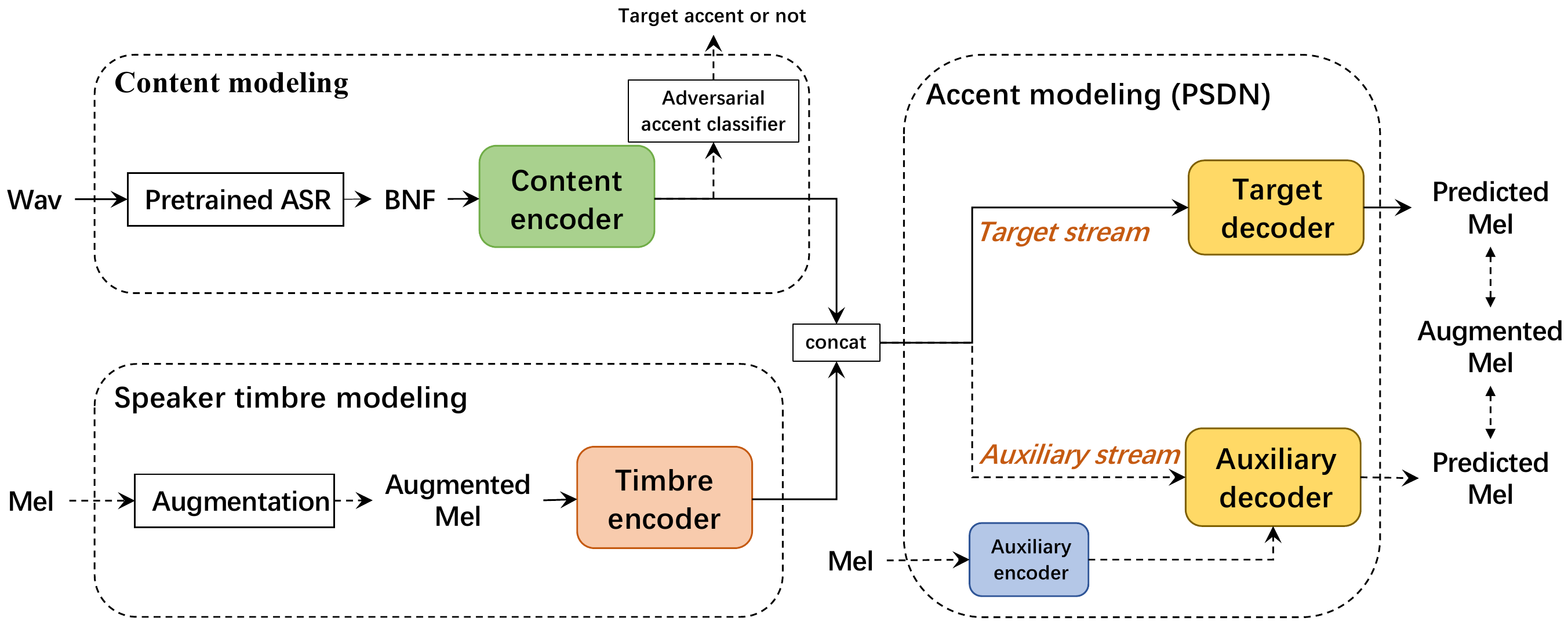}
    \caption{The workflow of the proposed method. The dotted lines are only enabled at the training stage.}
    \label{fig:flowchart}
\end{figure*}

\section{Related work}
Previous approaches to AC can be divided into two main categories, reference-based and reference-free. 
Referenced-based AC approaches, which build a VC model to transform reference utterances with the target accent from a source speaker to the target speakers, have been prevalent previously. 
Phonetic posteriogram (PPG) is often utilized to achieve the goal of VC~\cite{Zhao2018, zhao19f_interspeech}. 
But the basic VC model is only applicable to seen target speakers. To generalize to unseen speakers, the speaker encoder has been introduced~\cite{accentron} to extract a speaker embedding for unseen speakers at inference time.
To ease the requirement of reference at inference time, Li et al.~\cite{li2020improving} proposed a target-accent TTS system to generate reference speech at synthesis time.

To improve practical use, many reference-free methods have been proposed. Waris Quamer et al. ~\cite{waris2022interspeech} splits the AC task into two sub-tasks, pronunciation correction, and voice conversion. 
A translator is trained to map the BNF extracted from source-accent speech to that from target-accent speech.
The converted BNF is further transformed into Mel-spectrogram by a many-to-many VC model.
From another perspective, some previous works~\cite{zhao2021converting, nguyen22d_interspeech} build a VC model to generate synthetic parallel data and then train a sequence-to-sequence model to learn the mapping relationships of the parallel data. 
But these reference-free approaches are not applicable to unseen speakers.
Liu et al ~\cite{Liu2020icassp} have proposed a zero-shot system sharing the most similar advantage with our method. They propose to build a multi-speaker TTS model with a native English accent. An ASR model is trained to predict the output feature map of the encoder of the TTS model. 
Different from their method, our method achieves the goal of AC from the perspective of feature disentanglement. 

Siamese neural network (SNN) is an artificial neural network where two different inputs shared the same weight and the outputs are computed for similarity metric learning. It has been used for speech recognition~\cite{khorram2022contrastive} to leverage unlabeled acoustic data, and voice casting~\cite{8683178} to measure the proximity between the original and dubbed voice. 
Pseudo Siamese network (PSN) is a special case of SNN, where the inputs do not share weights and the structure of either stream is flexible. COMPOSE~\cite{Gao2020COMPOSECP} uses a cross-modal PSN for feature matching.
Xia et al.~\cite{10.1145/3404835.3462995} has proposed a PSN to generate labeled data for few-shot intents. To the best of our knowledge, the idea of SNN/PSN has not been used for feature disentanglement so far.
Gradient Reversal Layer (GRL) is an adversarial training trick,  which is originally proposed for unsupervised domain adaptation~\cite{ganin2015unsupervised} and widely adopted for feature disentanglement in various areas, such as music translation~\cite{mor2018autoencoderbased}, image-to-image translation~\cite{lee2018diverse} and text classification~\cite{liu2017adversarial}.

\section{Proposed method}

In the paper, we propose a zero-shot reference-free AC system. 
The workflow of the method is illustrated in Figure ~\ref{fig:flowchart}.
In this work, speaker timbre, content, and accent are modeled separately.
In content modeling, a content encoder is adopted to transform BNF into accent-free content representation, with the help of an adversarial accent classifier and following accent modeling. 
In accent modeling, PSDN is proposed to disentangle the source accent from the content representation and generate the target accent base on the content. PSDN is inspired by the idea of PSN and leverages the difference between two streams to achieve feature disentanglement.
In speaker timbre modeling, to better generalize to unseen speakers, a speaker augmentation method is proposed to extract an augmented speaker representation.

\subsection{Content modeling}
\label{ssec:content}

For input utterances, BNF is first extracted from a pre-trained ASR model, which is trained by Connectionist Temporal Classification (CTC) loss~\cite{graves2006connectionist}. To be specific, we adopt the output of the final hidden layer in the encoder of the ASR model as BNF. The extracted BNF is then fed into the content encoder, which is composed of multiple Conformer blocks~\cite{Gulati2020}, to obtain the content representation.

We expect the content representation to contain as little accent information as possible, otherwise, the source accent of input speech would affect the conversion performance. But we find that BNF contains fine-grained articulation information, including the accent.
To remove the accent information in the content representation, an adversarial accent classifier is first applied to the content representation to discriminate between the target accent and others, and Gradient Reversal Layer (GRL)~\cite{ganin2015unsupervised} is used to connect the content encoder and the accent classifier. To further achieve this goal, PSDN is proposed and presented in Section 3.3. The loss of content modeling is defined as:
\begin{align}
    F_{content} &= E_{content}(BNF)\\
    \hat{Y}_{accent} &= C_{accent}(GRL(F_{content}))\\
    L_{content} &= CE(Y_{accent}, \hat{Y}_{accent})
\end{align}
where $F_{content}$ denotes the output of the content encoder, $E_{content}$ denotes the content encoder, $C_{accent}$ denotes the accent classifier, $Y_{accent}$ and $\hat{Y}_{accent}$ denote the accent label and predicted accent, respectively. CE denotes Cross-entropy loss. $L_{content}$ denotes the loss for content modeling.

\subsection{Speaker timbre modeling}
\label{ssec:timbre}

To represent speaker timbre, Mel-spectrogram is fed into a timbre encoder to generate a global vector. We assume that accent is a time-varying feature, and thus the global timbre representation is irrelevant to accent.
During training, we propose a speaker timbre augmentation technique applied to the input of the timbre encoder, yielding an augmented Mel-spectrogram. To be specific, we adopt a pre-trained any-to-many VC model, which is able to change the speaker timbre while retaining the content and accent of the input. At the training stage, the speaker timbre of each utterance is randomly converted to some other speakers or remains unchanged. The main purpose is to increase the diversity of speaker timbre so as to improve the generalization ability of unseen speakers. This technique makes it possible to train our model on a very small quantity of target-accent speakers. Additionally, augmented Mel is used to compute loss (rightmost side in Figure~\ref{fig:flowchart}), ensuring speaker timbre is independently modeled and disentangled from the accent and content.

\subsection{ Accent modeling }
\label{ssec:accent}
Pseudo Siamese Disentanglement Network (PSDN) is proposed for accent modeling and plays two roles: first, reducing the source accent in content representation; second, generating the target accent given content and speaker timbre representations. 
PSDN consists of two streams, the target, and the auxiliary stream. The two streams have different architectures and receive different data during training. During inference, only the target stream is used. The details are shown on the right side of Figure~\ref{fig:flowchart}. The differences between the two streams can be summarized into two aspects: 

(1) Received information. The auxiliary stream receives more information as input than the target stream. To be specific, the auxiliary stream contains an extra encoder to receive Mel-spectrogram without augmentation, which is rich in accent-related information.

(2) The difficulty of the training task. The data flowing through the two streams are different. Here we use an example for a better explanation. Assuming the British accent is the target accent, and our training data consist of British-accent and other-accent data. In training, British-accent data flow through both the target and the auxiliary stream, while other-accent data only flow through the auxiliary stream. 
The task of the auxiliary stream is to reconstruct the accent of all speakers, which is obviously more difficult than that of the target stream. Therefore the model may prioritize fulfilling the need of the auxiliary stream, which is to obtain accent-related information as easily as possible.

 We take advantage of the differences between the two streams to realize the disentanglement of the accent. For the auxiliary stream, there are two potential sources of accent information: the content representation and the auxiliary encoder. Since an adversarial accent classifier has been applied to the content representation, it is intuitively easier for the auxiliary stream to learn the accent information from the auxiliary encoder rather than the content representation. Therefore, the accent information in the content representation is forced to decrease. Due to the lack of accent information in the content representation, thus the target stream will learn to model the target accent given the content. Additionally, there is an extra merit of PSDN. Since we often have a large amount of other-accent data, the auxiliary stream amounts to an exit for these data so that the speaker timbre and content modules have the opportunity to learn from more data.

We compute the $L1$ loss between the output of each stream with the augmented Mel-spectrogram. The loss of accent modeling is defined as :
\begin{align}
    \hat{Y}_{target} &= S_{target}(F_{content}, F_{timbre})\\
    \hat{Y}_{aux} &= S_{aux}(F_{content}, F_{timbre}, F_{aux})\\
     L_{accent} &= \underbrace{\lVert Y_{aug\_mel}-\hat{Y}_{target}\rVert_{1}}_{\text{For target-accent data}} + 
    \underbrace{\lVert Y_{aug\_mel}-\hat{Y}_{aux}\rVert_{1}}_{\text{For all data}}
\end{align}
where $\hat{Y}_{target}$ and $\hat{Y}_{aux}$ denote the outputs from the target and auxiliary stream, respectively. $S_{target}$ and $S_{aux}$ denote the target and the auxiliary stream, respectively. $F_{timbre}$ and $F_{aux}$ denote the outputs from the timbre encoder and auxiliary encoder, respectively. $Y_{aug\_mel}$ denotes the augmented Mel-spectrogram. $L_{accent}$ is the loss for accent modeling. To summarize, the loss used to train the proposed system is defined as:
\begin{equation}
    L = L_{content} + L_{accent}
\end{equation}

\begin{table}[h!]
\centering
\caption{Training/validation and test data. The training/validation set consists of the target-accent and other-accent data.}
\begin{tabular}{ c c} 
 \hline
\multicolumn{2}{c}{Foreign-to-native (British)}\\
 \hline
 Target-accent & 1 speaker, 10H (Inhouse)\\ 
 Other-accent & 1122 speakers, 251H (LibriTTS~\cite{zen2019libritts}) \\ 
 Test & 4 foreign-accent speakers (ARCTIC-L2)\\ 
 \hline
 \multicolumn{2}{c}{Native-to-foreign (Indian)}\\
  \hline
 Target-accent & 260 speakers, 39H (Inhouse) \\ 
 Other-accent & 1122 speakers, 251H (LibriTTS) \\ 
 Test & 4 American native speakers (Inhouse) \\ 
 \hline
\end{tabular}
\label{table:data}
\end{table}

\section{Experimental setup}
\label{sec:exp}

\subsection{Data}
We conduct experiments on both foreign-to-native and native-to-foreign conversion, using British and Indian accents as the target accent, respectively. The datasets for the two experiments are summarized in Table 1. 
Since LibriTTS has a great diversity of speakers, we use train-other-500 of LibriTTS as other-accent data for both experiments. 
For foreign-to-native conversion (British), the target-accent data contains only 1 British-accent speaker. Although LibriTTS also contains some unlabeled British speakers, there are differences between individuals and thus we can treat them as other-accent. The test set is non-native English speech from L2-ARCTIC~\cite{zhao2018l2}.
For native-to-foreign conversion (Indian), the target-accent data contains 260 Indian-accent speakers. Different from the above British case, we treat accent as a group attribute here. The test set is some native American speech from inhouse dataset. For each experiment, we randomly select 5\% of utterances for each speaker as the validation set. 

\subsection{Model architecture details}
\label{ssec:exp:modelar}

The content encoder is composed of a 3-layer Conformer~\cite{Gulati2020} with a hidden size of 512.
The accent classifier is composed of a bidirectional LSTM~\cite{hochreiter1997long} with a hidden size of 256 for each direction, and 4 convolutional residual blocks with a channel size of 256. Each convolutional residual block contains 2 convolution layers followed by LeakyReLU activation. The down-sampling rates of the residual blocks are 4, 2, 2, and 2, respectively. The pre-trained ASR model for BNF extraction follows the architecture of~\cite{Gulati2020} and is trained on LibriSpeech~\cite{panayotov2015librispeech} with CTC loss computed among 5013 subwords~\cite{zenkel2018subword} in English. 

The architecture of the timbre encoder follows the architecture of~\cite{wang2018style}, which is composed of a reference encoder and a style token layer. The reference encoder contains a 2-D convolution stack and a GRU~\cite{bahdanau2014neural} layer. The convolution stack contains 4 Conv-BatchNorm~\cite{ioffe2015batch}-ReLU blocks, where the kernel size and stride and (3, 3) and (2, 2) for each block, and the channel size is [16, 32, 64, 128]. The hidden size of the GRU layer is 256. For the style token layer, 20 tokens with 256 channels are used. 
The pre-trained any-to-many VC model used for speaker augmentation follows the architecture of~\cite{liu2021delightfultts}. The model is trained on LibriTTS and VCTK~\cite{veaux2017cstr}, among which speakers with less than 100 sentences are excluded. 

\begin{table}[h!]
\centering
\caption{Mean opinion score (MOS) for naturalness and accentedness with a 95\% confidence interval on two target accents.}
\resizebox{\linewidth}{!}{
\begin{tabular}{ c| c c| c c} 
 \hline
Systems & \multicolumn{2}{c}{Naturalness$\uparrow$} & \multicolumn{2}{|c}{Accentedness$\uparrow$}\\ 
   \hline
 & British & Indian & British & Indian \\
  \hline
 Input  & 3.28$\pm$0.14 & \textbf{4.32$\pm$0.09} & 2.05$\pm$0.10 & 1.12$\pm$0.07\\ 
 \hline
 Baseline & 3.41$\pm$0.13 & 3.97$\pm$0.12 & 2.57$\pm$0.13 & 1.31$\pm$0.10\\ 
 Proposed & \textbf{3.45$\pm$0.10} & 3.95$\pm$0.08 & \textbf{3.35$\pm$0.12} & \textbf{3.82$\pm$0.12}\\ 
 \hline
\end{tabular}
}
\label{table:mos}
\end{table}

\begin{table}[h!]
\centering
\caption{Mean opinion score (MOS) for speaker timbre similarity with a 95\% confidence interval on two target accents.}
\begin{tabular}{ c| c c} 
 \hline
 Systems &  \multicolumn{2}{c}{Speaker similarity$\uparrow$ }\\ 
 \hline
  & British & Indian\\
  \hline
 Proposed & \textbf{3.13$\pm$0.07} & \textbf{3.81$\pm$0.15}\\ 
 w/o speaker augmentation & 1.80$\pm${0.11} & 1.90$\pm${0.10}\\ 
 \hline
\end{tabular}
\label{table:smos}
\end{table}

Both the target decoder and auxiliary decoder are a 3-layer Conformer with a hidden size of 512. The auxiliary encoder is composed of 4 convolutional residual blocks with hidden sizes of 128, 128, 128, and 16. Each convolutional residual block contains 2 convolution layers followed by the LeakyReLU activation. During conversion, a pre-trained universal TFGAN vocoder~\cite{tian2020tfgan} is used to transform the predicted Mel-spectrogram to the waveform.

The proposed model is trained for 220k steps with the Adam optimizer~\cite{kingma2014adam} using a batch size of 32 and a constant learning rate of 0.001. The training process takes 2 days on 4 Tesla V100 GPUs. Each mini-batch contains an equal number of target-accent and non-target-accent speech data. Logarithmic 80-dimensional Mel-spectrograms with a hop size of 240ms are extracted from 24000Hz waveforms. The $\lambda$ of GRL is $5e^{-3}$. 

\subsection{Baseline and ablation study}
\label{ssec:exp:eval}
In the baseline method, we remove the proposed PSDN and keep the accent classifier with GRL for feature disentanglement, since GRL proved to be effective for feature disentanglement in many areas~\cite{mor2018autoencoderbased,lee2018diverse,liu2017adversarial}.
To be specific, we replace PSDN with two separate decoders of identical architecture, one for the target accent and the other for other accents. The other parts remain the same as the proposed method.

We also conduct an ablation study on the proposed speaker augmentation method to verify its effectiveness. We directly remove it during training so that the augmented Mel-spectrogram is replaced by the original one.

\section{Results}
\label{ssec:exp:results}

To evaluate the proposed system, three perceptual listening tests are conducted, including audio naturalness, accentedness, and speaker similarity. A standard 5-point scale mean opinion score (MOS) is used for all the tests.
For either native-to-foreign (Indian) or foreign-to-native (British) experiments, there is a total of 20 test utterances from 4 speakers consisting of 2 females and 2 males. All speakers in the test set of each experiment are unseen during training. 10 participants proficient in English are recruited for the listening tests. 

\subsection{Naturalness}
All participants are asked to rate the acoustic quality of each utterance from the baseline and proposed method. Following~\cite{waris2022interspeech}, the input utterances are also evaluated for comparison. The results are shown in Table~\ref{table:mos}. 
In general, both the baseline and proposed systems obtain comparable MOS to that of the input, demonstrating that our system does well in the preservation of audio naturalness. It is reasonable that the baseline shares a similar MOS with the proposed system because either system uses the target stream for inference, which is consistently trained on the target-accent data.

In particular, the proposed method yields even higher MOS than that of the input in the foreign-to-native (British) experiment.
The main reason is that the non-native test samples for the British accent are from L2-ARCTIC and have obvious background noise. The proposed method lets other-accent data of uneven quality only flow through the auxiliary stream, guaranteeing the target stream is trained with relatively high-quality speech. This illustrates that the proposed system can improve audio quality to a certain extent when the input is noisy.

\subsection{Accentedness}
For each utterance from the baseline and proposed method, all participants are asked to rate the degree of proximity to the target accent. During the evaluation process, some target-accent utterances from the training set are provided to participants for comparison and reference.
The results are shown in Table~\ref{table:mos}. It can be seen that the proposed system outperforms the input and the baseline by a large margin on both foreign-to-native and native-to-foreign conversion. This demonstrates that the proposed PSDN can effectively disentangle the source accent from the content, while the baseline system only with GRL, fails to disentangle the accent from the content representation and tends to retain the source accent of the input. We find that the accentedness variance among all systems in foreign-to-native (British) is smaller than that of native-to-foreign (Indian). The reason is probably that the British accent is more difficult to distinguish than the Indian accent.

\subsection{Speaker similarity}
The purpose of the listening test is to verify the effectiveness of the proposed speaker augmentation method. In this test, each sample from the proposed and the ablation system is paired with the corresponding input speech, then participants are asked to listen to a pair of audio samples and rate the speaker timbre similarity. A higher score represents higher identity similarity.
Results are shown in Table~\ref{table:smos}. It can be seen that the proposed system yields a much higher similarity score than the system without speaker timbre augmentation on both foreign-to-native and native-to-foreign conversion. 
In particular, for the British case, the training data only contain one target-accent speaker. The proposed speaker augmentation method can increase the number of target-accent speakers and alleviate the problem of a lack of target-accent speakers. We can conclude that the proposed speaker augmentation technique can improve the generalization ability on unseen speakers, even with very limited target-accent speakers in training data.

\section{Conclusion}
\label{sec:conclusion}

In this work, we propose a zero-shot reference-free accent conversion method requiring only non-parallel data at the training stage, which is applicable to two-way conversion. PSDN is proposed to solve the AC problem from the perspective of feature disentanglement. 
Experimental results show that the proposed method is able to convert unseen speakers' utterances into the target accent with high naturalness and accentedness, on both foreign-to-native and native-to-foreign conversion. The speaker similarity is also largely improved by the proposed speaker augmentation method.
In the future, the ability to retain speaker identity will be improved for unseen speakers through further research. Additionally, the proposed PSDN is widely applicable to various tasks on feature disentanglement.

\bibliographystyle{IEEEtran}
\bibliography{mybib}

\end{document}